\documentclass[%
 reprint,showpacs,
 amsmath,amssymb,
 aps,prl
]{revtex4-1}

\usepackage{graphicx}  
\usepackage{dcolumn}   
\usepackage{bm}        
\usepackage{amssymb}   

\usepackage[normalem]{ulem}
\usepackage[usenames,dvipsnames]{xcolor}
\usepackage{color}

\begin{document}

\title{Structual Vulnerability of the Nematode Worm Neural Graph}

\author{Michelle Rudolph-Lilith}
\email[Electronic address: ]{michelle.rudolph@unic.cnrs-gif.fr}
\author{Alain Destexhe}
\author{Lyle E. Muller}

\affiliation{\mbox{Unit\'e de Neurosciences, Information et Complexit\'e (UNIC), CNRS, 1 Ave de la Terrasse, 91198 Gif-sur-Yvette, France}}

\date{\today}

\begin{abstract}
The number of connected components and the size of the largest connected component are studied under node and edge removal in the connectivity graph of the {\it C. elegans} nervous system. By studying the two subgraphs -- the directed graph of chemical synapses and the undirected graph of electrical junctions -- we observe that adding a small number of undirected edges dramatically reduces the number of components in the complete graph. Under random node and edge removal, the {\it C. elegans} graph displays a remarkable structural robustness. We then compare these results with the vulnerability of a number of canonical graph models. 
\end{abstract}

\maketitle

Many natural and artificial systems can be described in terms of graphs, assemblies of nodes pairwise related through edges \cite{GraphTheory}. One subject receiving attention in recent literature is the study of robustness, or resilience, of graphs subjected to the removal of either nodes or edges \cite{Vulnerability1,Vulnerability2,Vulnerability4}. In technical systems, such as power grids or the internet, a low vulnerability (high resilience) to malfunctioning nodes or the interruption of links is essential to ensure a continuous and stable service \cite{Vulnerability3,Attacks1}.  Many of these studies have focused on targeted or guided attacks, in which nodes/edges are removed depending on their degree \cite{Attacks1,Attacks2}, betweenness \cite{Attacks1,Attacks3} or range \cite{Vulnerability4}, providing valuable insights into the structural prerequisites ensuring high resilience. These studies also suggest that the assessment of various graph-theoretical measures as a function of node/edge removal can be used as a valuable tool to probe the structural characteristics of graphs. 

We posit here that such a feature of resilience should be desirable for biological neuronal networks. Upon removal of nodes or edges, a graph will typically decompose into an increasing number of smaller connected components -- assemblies of nodes no longer mutually interlinked. The vulnerability of the graph is proportional to the rate of its decomposition under node and edge removal. Additionally, graph robustness can be measured by the relative size of its largest (giant) connected component. Graphs with a connected component comprised of the majority of nodes can be considered to have low vulnerability, as this dominant component is more likely to contain the assembly of nodes/edges crucial to the graph's function.

In this letter, we use the number of connected components $N_{cc}$ and the size (defined as the number of nodes) of the largest connected component $S_{gcc}$ \cite{GraphTheory}, to explore the structural vulnerability of the nematode {\it Caenorhabditis elegans} neural network with respect to uniform-random node and edge removal. In contrast to most previous studies, which applied a wide range of graph-theoretical measures to the undirected version of the {\it C. elegans} graph \cite{Celegans1}, we study its original directed version, with only self-loops (accounting for about 0.1\% of edges) excluded. In directed graphs, two nodes are strongly connected if a directed path exists through which both nodes can reach each other. A set of strongly connected nodes forms a strongly connected component, which we will refer to as the connected component. In the first part of the study, both measures are applied to the individual subgraphs -- the undirected graph of electric junctions and the directed graph of chemical synapses. In the second part, we compare the obtained results with several canonical graph models. Throughout the study, 10,000 random realizations at each data point were drawn to ensure statistical validity. The numerical analysis was performed using custom software, which, along with all analysis protocols used, is available for download \cite{CYDYNS}. 

The neuronal graph of the hermaphrodite worm is, to date, the only naturally occurring neural graph for which an almost complete wiring diagram has been made available \cite{Celegans1, Celegans2}. The complete graph (CG) is sparse (connectance $Co = N_E/N_E^{max}$~=~0.038; \cite{GraphTheory}) and consists of $N_N$~=~279 nodes with a total of $N_E$~=~2,990 edges and an average node in/out-degree of $\langle a_i \rangle$~=~10.72. From these, 514 edges are electric junctions (EJ), forming an undirected subgraph ($Co$~=~0.013) with $\langle a_i \rangle$~=~3.68, and 2,194 are chemical synapses (CS) with $\langle a_i \rangle$~=~7.86 forming a directed subgraph ($Co$~=~0.028). The complete {\it C. elegans} graph has 6 connected components, with the largest component comprised of 274 nodes. The CS (EJ) subgraph has 42 (29) connected components with 237 (248) nodes in its largest component. Importantly, although the CS subgraph contributes more than 73\% of edges to the complete graph, it has 7 times more connected components than the complete graph. This number is dramatically reduced by the addition of a comparably small number (27\%) of undirected edges, suggesting a complementary or synergistic role of both subgraphs.

\begin{figure}
\centerline{\includegraphics[width=\columnwidth]{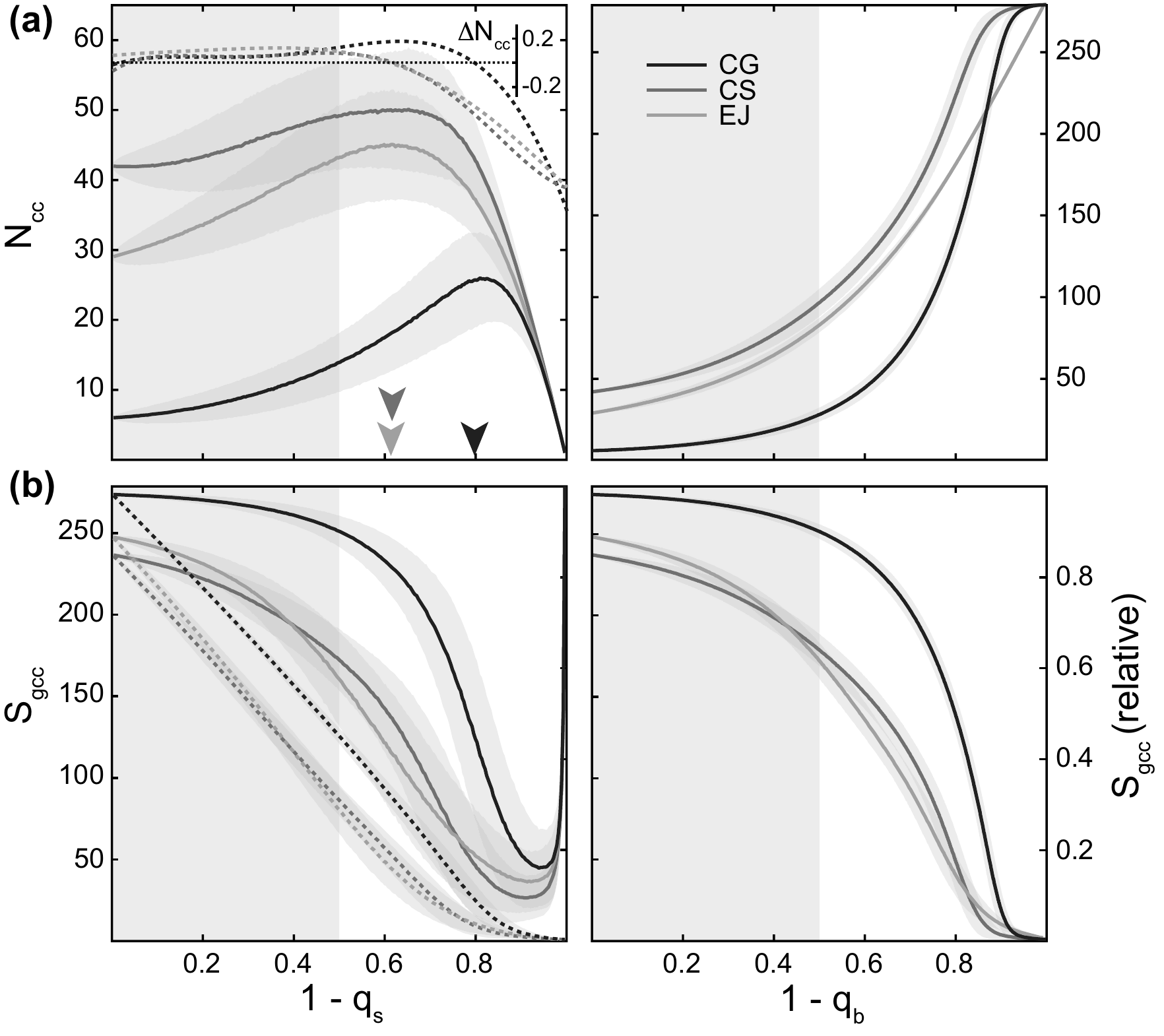}}
\caption{(a) $N_{cc}$, (b) $S_{gcc}$ (dotted; tic marks left), and relative $S_{gcc}$ (solid; tic marks far right) as functions of $1-q_{\{s,b\}}$ for the {\it C. elegans} graph, and its chemical synapse (CS) and electric junction (EJ) subgraphs (mean $\pm$ SD). Dotted lines in (a) show polynomial-fitted numerical derivatives (dotted), with arrowheads indicating PLD (see text).}
\label{Fig_C_elegans}
\end{figure}

This remarkable synergistic tendency is also observed when nodes and edges are removed from the graphs (Fig.\ref{Fig_C_elegans}a). Denoting with $q_{\{s,b\}}$ the fraction of occupied nodes/edges, both the CS and EJ subgraphs display a large $N_{cc}$, which increases modestly with node/edge removal, despite the fact that the number of edges decreases almost linearly with the number of removed nodes/edges (not shown). The complete graph is characterized by a significantly smaller $N_{cc}$. Even after removal of 50\% of nodes/edges, $N_{cc}$ remains comparably small ($N_{cc}$~=~13.8~$\pm$~2.3 for node removal, 28.1~$\pm$~2.4 for edge removal), a surprising result given the sparseness of the {\it C.elegans} graph. As removing nodes will naturally yield $N_{cc}$~=~0 for $q_s$~=~0, for some intermediate value of $q_s$ a {\it point of largest decomposition} (PLD) may be reached, defined as the (global) maximum of $N_{cc}$. A numerical difference analysis (Fig.\ref{Fig_C_elegans}a, dotted lines) shows that the PLD for both subgraphs is almost identical ($q_s \sim$~0.4, i.e. for 60\% of removed nodes), whereas it shifts significantly to lower $q_s$ (80\% removed nodes) for the complete graph, providing further support for the complimentary combination of the individual subgraphs.

A similar resilience is observed with $S_{gcc}$ under node/edge removal. As expected, the largest component in the complete graph is always bigger than that of its subgraphs, and $S_{gcc}$ decreases quasilinearly with a slope of about -1.11 (-1.07 for CS, -1.18 for EJ; Fig.\ref{Fig_C_elegans}b, dashed). This linear decrease is somewhat surprising and more expected from unstructured graphs (e.g.~Erd\H{o}s-R\'{e}nyi graph) as it suggests a homogeneous and isotropic connectivity pattern. This gradual linear decrease translates directly into an only modest decrease of the relative size (fraction of remaining nodes) of the largest component as a function of $q_s$ (Fig.\ref{Fig_C_elegans}b, solid). After removal of 50\% of the nodes, the largest component contains still more than 90.3\% of the remaining nodes in the graph (98\% for $q_s$~=~1). Edge removal displays similar behavior, with 90.1\% of the nodes remaining in the largest component for $q_b$~=~0.5, further corroborating the low structural vulnerability of the {\it C. elegans} neural graph. 

In order to further explore structural characteristics of the {\it C. elegans} graph, we applied the above analysis to a number of canonical graph models constructed to yield the same $N_N$ and about the same $N_E$ as the complete {\it C. elegans} graph as well as its CS and EJ subgraphs:

\begin{figure}[b]
\centerline{\includegraphics[width=\columnwidth]{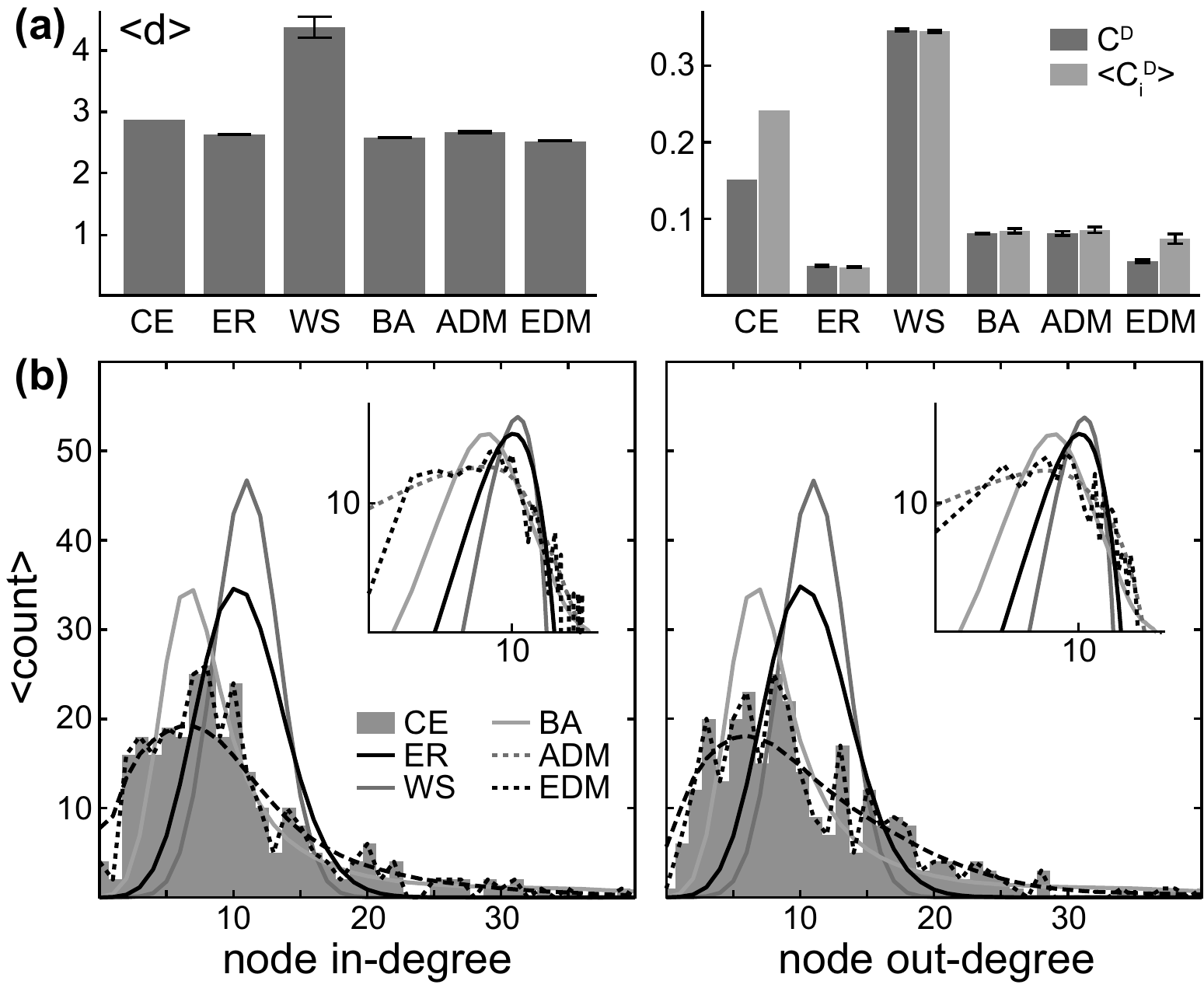}}
\caption{Average geodesic graph distance $\langle d \rangle$, global and average node clustering coefficient ($C^D$ and $\langle C_i^D \rangle$), and node in/out-degree distributions for the {\it C.elegans} (CE) and a number of canonical graph models (see text). The insets show log-log plots of the degree distributions.}
\label{Fig_Small_Worldiness}
\end{figure}

{\it Erd\H{o}s-R\'{e}nyi model} (ER) \cite{ERmodel}. We employed the classical directed and undirected ER model, which chooses uniform randomly from the ensemble of all graphs with a given number of nodes and edges. 

{\it Watts-Strogatz model} (WS) \cite{WSmodel}. In this classical example of small-world graphs, average node degrees were set to 11 (yielding $N_E$~=~3,069) for the directed model matching CG, 8 ($N_E$~=~2,232) for CS, and 2 ($N_E$~=~558) for EJ. A rewiring probability $p^{w}$~=~0.01 was chosen for all models to ensure high clustering coefficient and small average geodesic graph distance. To construct directed WS graphs, we first constructed appropriate undirected WS graphs after which the direction of each edge was randomized with a probability of 0.5.

{\it Barab\'{a}si-Albert model} (BA) \cite{Vulnerability2}. For random scale-free graphs, we used the BA model of preferential attachment. The initial number of nodes were 11, 8 and 2 for the CG, CS and EJ, respectively. Initial node degrees and increments were fixed to 11 (CG), 8 (CS) and 2 (EJ), yielding $N_E$~=~3,003 (CG), 2,196 (CS) and 555 (EJ). Directed versions were obtained from appropriate undirected BA graphs and randomization of edge direction.

{\it Approximate degree-matched random model} (ADM). In order to discern to which extent the resilient behavior of the {\it C. elegans} graph depends on its node degrees, we constructed a model which approximates uniform sampling from the collection of all graphs with a given degree distribution. The adjacency matrix $a_{ij}$ was defined as $a_{ij}$~=~1 if $p < \frac{1}{a} a_i^{out} a_j^{in}$, where $a_i^{\{in,out\}}$ denote the node in/out-degrees of the target graph and $a = \sum_i a_i^{\{in,out\}}$, and $a_{ij}$~=~0 otherwise. A random number $p \in [0,1)$ was chosen for each randomly selected index pair $(i,j)$ with $i,j \in [0, N_N)$ until $N_E$ edges were generated. 

{\it Exact degree-matched Erd\H{o}s}-R\'{e}nyi model (EDM). Finally, a sophisticated model introduced by Kim and colleagues \cite{EDMmodel} was used which provides a biased sampling of the set of graphs with specific node degrees.

We first compared the average geodesic graph distance $\langle d \rangle$ \cite{GraphTheory}, the global clustering coefficient $C^D$, the average node clustering coefficient $\langle C_i^D \rangle$ \cite{ClusteringCoefficient} and node degree distributions in the various models. As expected, none of the canonical models (ER, WS, BA) reproduced the node degree distribution observed in the {\it C. elegans} graph (Fig.\ref{Fig_Small_Worldiness}b). The latter was, however, well approximated by the ADM model. Somewhat surprisingly, the low value of $\langle d \rangle$ was shared among all models, with the exception of WS (Fig.\ref{Fig_Small_Worldiness}a). Although the WS model allows adjustment of this value through a change of its rewiring probability $p^{w}$, a lower value of $\langle d \rangle$ can only be achieved on the expense of a reduction in its small-worldness index \cite{SWIndex}, defined here for directed graphs as
$$
S^{D} = \frac{C^D}{C^D_{\mbox{\tiny EDM}}} \frac{\langle d \rangle_{\mbox{\tiny EDM}}}{\langle d \rangle}
$$
where $C^D_{\mbox{\tiny EDM}}$ and $\langle d \rangle_{\mbox{\tiny EDM}}$ are the corresponding values for the EDM model. Choosing $p^{w}$~=~0.2 yields $\langle d \rangle$~=~2.87 and $C^D$~=~0.19, values comparable to that observed in the {\it C. elegans} graph. However, $S^D$ will decrease from 4.47~$\pm$~0.03 for $p^{w}$~=~0.01 to 3.73~$\pm$~0.05 for $p^{w}$~=~0.2 ($S^D$~=~2.98 for {\it C. elegans}, 1.77 for BA, 0.83 for ER). In this study, we choose to explore a WS model in which $S^D$ is approximately maximal. Our findings suggest that a characterization of the {\it C. elegans} neural graph as small-world or scale-free should receive caution \cite{MythPaper}, although further study is necessary to support this conclusion. 

\begin{figure}
\centerline{\includegraphics[width=\columnwidth]{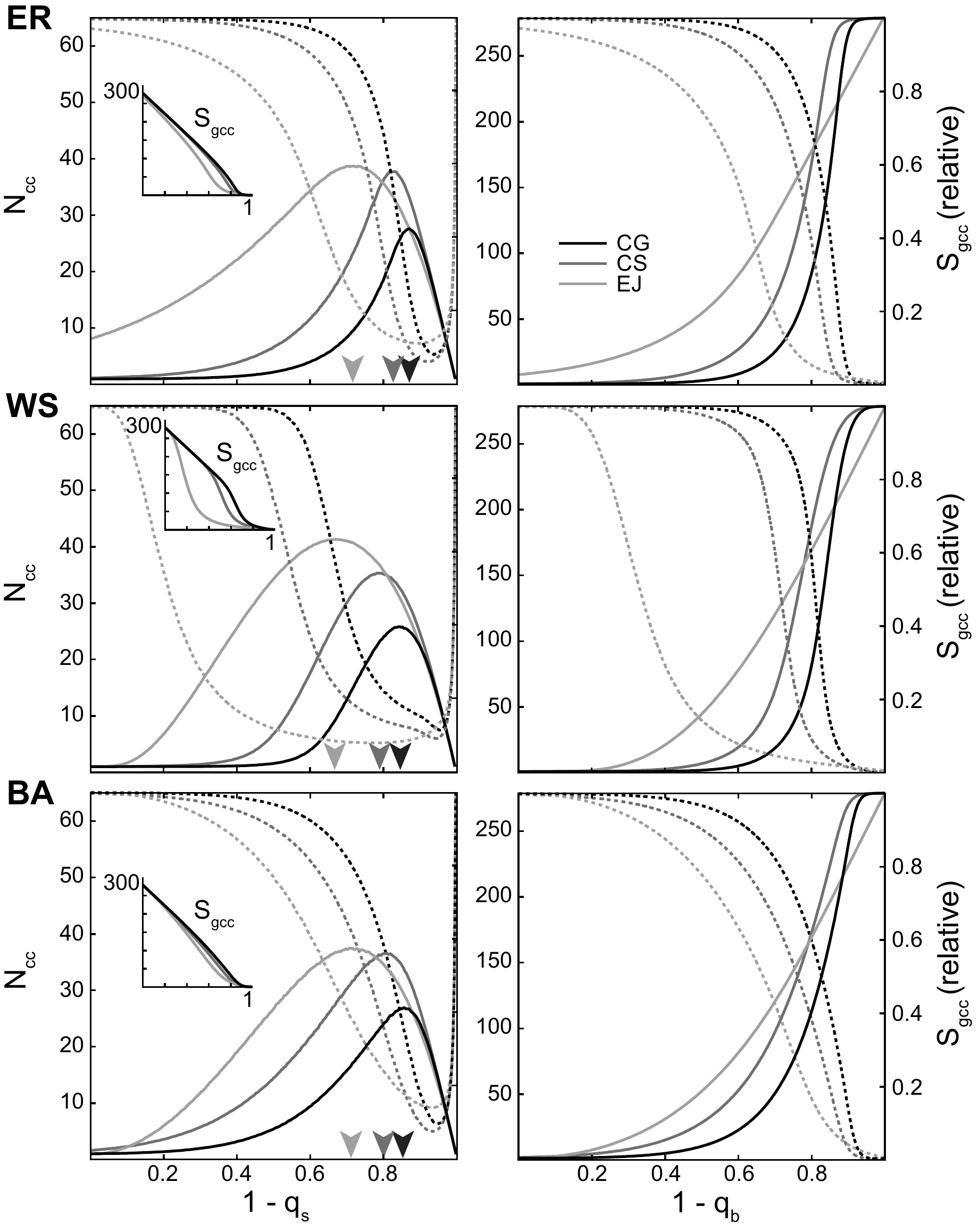}}
\caption{$N_{cc}$ (solid; tic marks left) and relative $S_{gcc}$ (dashed; tic marks far right) for node (left) and edge (right) removal in canonical graph models (see text). Models corresponding to the complete graph and the CS/EJ subgraphs are shown as in Fig.~\ref{Fig_C_elegans}. Insets display the absolute size of the the largest connected component. Arrowheads indicate PLD.}
\label{Fig_ER_WS_BA}
\end{figure}

\begin{figure}
\centerline{\includegraphics[width=\columnwidth]{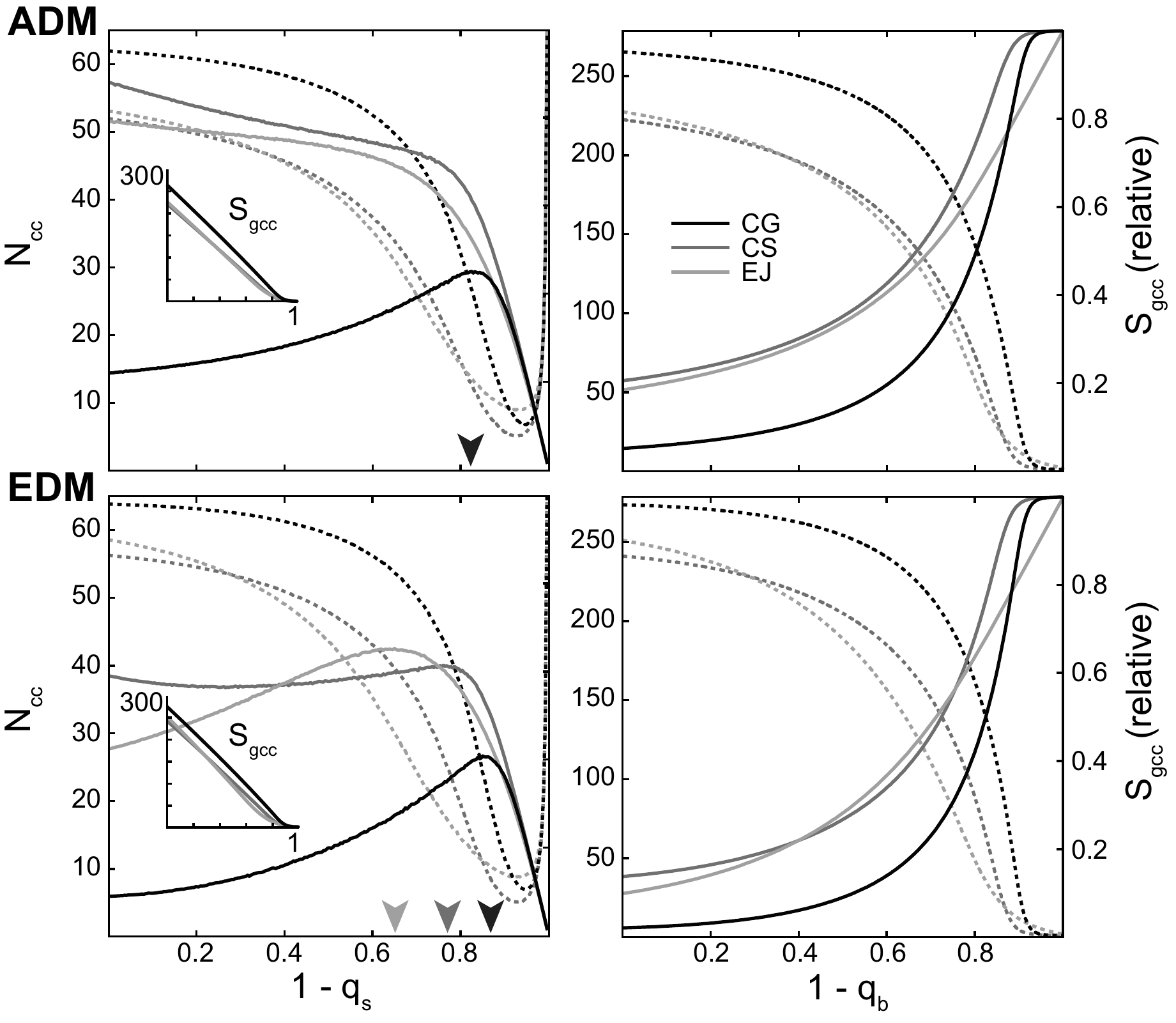}}
\caption{$N_{cc}$ (solid; tic marks left) and relative $S_{gcc}$ (dashed; tic marks far right) for node (left) and edge (right) removal in the ADM and EDM model (cf. Fig.~\ref{Fig_ER_WS_BA}). }
\label{Fig_ADM_EDM}
\end{figure}

Further support for this cautious assessment is provided by the vulnerability analysis. None of the canonical graph models was capable of reproducing the synergy between the subgraphs which gives rise to the high resilience observed in the complete {\it C. elegans} graph (Fig.\ref{Fig_ER_WS_BA}). The $N_{cc}$ for large $q_{\{s,b\}}$ was significantly lower than in {\it C. elegans} by construction, and increased rapidly after reaching a critical node/edge occupation probability $q_{\{s,b\}} \sim$~0.4 (ER, WS) or 0.6 (BA) for CG. Such a critical behavior was not observed in {\it C. elegans}. Furthermore, the PLD in these canonical models displayed a clear dependency on the average node degree, with the EJ subgraph showing the lowest PLD. Interestingly, both ADM and EDM captured, for CG, the dependence of $N_{cc}$ on $q_{\{s,b\}}$ seen in {\it C. elegans} (Fig.\ref{Fig_ADM_EDM}). While both subgraphs of ADM deviated significantly from {\it C. elegans}, only the CS of EDM differed, with EJ capturing the behavior of the biological graph under node/edge removal. Taken together, these observations suggest that while the degree-matched graphs provide the best model for the structural decomposition of {\it C. elegans}, there remains a small but significant structural component which is not explained by a degree-matched random connectivity alone.  

Differences between {\it C. elegans} and canonical graph models were also prominent when considering $S_{gcc}$. In neither ER, WS or BA a quasilinear dependency of $S_{gcc}$ was observed, but a clear supralinear decrease, most significant in the WS model (Fig.~\ref{Fig_ER_WS_BA}, insets). For the ER model this finding is particularly surprising, as its isotropy and homogeneity should ensure a decrease of $S_{gcc}$ proportional to $q_s$. Deviations from this expectation must be attributed to the small size of the graph and its low connectance. This behavior of $S_{gcc}$ is reflected in its relative measure. Here, after reaching a critical occupation probability, the relative $S_{gcc}$ decreases rapidly, and the CS and EJ subgraphs show significant quantitative differences, a finding not observed in {\it C. elegans}. Finally, the similarity in the dependency of $S_{gcc}$ on $q_{\{s,b\}}$ in the ADM and, especially, the EDM when compared to the corresponding biological graphs strongly supports the conclusion that the structural resilience of the {\it C. elegans} neural graph is driven by its degree distribution.

In summary, this letter addressed the structural vulnerability of the nematode worm neural graph by studying the behavior of the number of connected components and the size of the largest connected component as function of uniform random removal of its nodes and edges. We observed a strong structural resilience in the {\it C. elegans} graph as well as its directed chemical synapse and undirected electric junction subgraphs, under both node and edge removal. This is surprising given the low connectance of these graphs. Furthermore, the robustness of the residual largest component to perturbations in connectivity suggests that the {\it C. elegans} neural system may remain functional even after removal of a significant fraction of its neurons or synapses, as expected from naturally systems exposed to detrimental external influences. 

Our study showed further that the chemical synapse and electric junction subgraphs of the {\it C. elegans} display a remarkably similar behavior in both measures, despite significant differences in connectance, and that the complete neural graph is more than a simple combination of the two subgraphs. This observation was not shared in the investigated canonical graph models. Only models of degree-matched random graphs reproduced the various structural features of {\it C. elegans} studied here. This suggests that the structural vulnerability of the nematode neural graph is critically determined by its node degree distribution, and less by specific connectivity properties such as scale-free- or small-world-ness. Based on the results of this study we cautiously conclude that the clustering coefficient and geodesic path length alone are not sufficient for a characterization of the {\it C. elegans} neural and other real-world graphs, and that with respect to the measures of structural vulnerability considered here, the {\it C. elegans} graph is best modeled by an (exact) degree-matched random graph.

Our study does not address in detail whether the synergistic behavior of the combination of an undirected and directed subgraph can be reproduced by a simple increase in the number of edges, or whether it is the product of a complementary mixing of an undirected and directed graph. The results presented here suggest the latter, as the addition of only a small number of undirected edges significantly reduces the structural vulnerability of the  {\it C. elegans} neural graph. By taking into account the degree distributions of individual subgraphs, a specific mechanism 
for synergistic combination of directed and undirected graphs in constructing biological neural graphs could be conceived, and will be considered in future work.

\end{document}